\documentclass[aps,showpacs,preprintnumbers,amsmath, amssymb]{revtex4}

\usepackage{float}
\usepackage{graphics,epsfig}
\usepackage{graphicx}
\usepackage{dcolumn}
\usepackage{bm}

\begin{document}
\baselineskip=0.8 cm
\title{Various types of phase transitions in the AdS soliton background}
\author{Yan Peng } \affiliation{Department of Physics, Fudan University, 200433 Shanghai, China}
\author{Qiyuan Pan, Bin Wang}   \affiliation{INPAC and Department of Physics, Shanghai Jiao Tong University, 200240 Shanghai, China}

\vspace*{0.2cm}
\begin{abstract}
\baselineskip=0.6 cm
\begin{center}
{\bf Abstract}
\end{center}

We study the basic holographic insulator and superconductor phase
transition in the AdS soliton background by generalizing the
spontaneous breaking of a global U(1) symmetry to occur via
St$\ddot{u}$ckelberg mechanism. We  construct the soliton solutions
with backreaction and examine the effects of the backreaction on the
condensation of the scalar hair in the generalized
St$\ddot{u}$ckelberg Lagrangian. We disclose rich physics in various
phase transitions. In addition to the AdS soliton configuration,  we
also examine the property of the phase transition in the AdS black
hole background.
\end{abstract}

\pacs{11.25.Tq, 04.70.Bw, 74.20.-z}\maketitle
\vspace*{0.2cm}

\section{Introduction}

The AdS/CFT correspondence \cite{Maldacena,S.S.Gubser-1,E.Witten}
has provided a framework to describe the strongly coupled field
theories in a weakly coupled gravitational system. In recent years,
this correspondence has been applied to study the holographic model
of superconductors, which is constructed by a gravitational theory
of a Maxwell field coupled to a charged complex scalar field
\cite{liu4,liu5,liu6}. It has been shown that the bulk AdS black
hole becomes unstable and scalar hair condenses below a critical
temperature. In the boundary dual CFT, many properties shared by the
superconductor have been exhibited. The instability of the bulk
black hole corresponds to a second order phase transition from
normal state to superconducting state which brings the spontaneous
U(1) symmetry breaking. The gravity models with the property of the
holographic superconductor have attracted considerable interest for
their potential applications to the condensed matter physics, see
for example \cite{liu7}-\cite{liu33}.

Considering that the spontaneous breaking of a global U(1) symmetry
in the basic holographic superconductor model occurs via the
St$\ddot{u}$ckelberg mechanism, recently it was disclosed that
besides the second order phase transition, a fairly wide class of
phase transitions can be allowed to happen \cite{pan26}. The
St$\ddot{u}$ckelberg mechanism allows tuning the order of the phase
transition which can accommodate the first order phase transition to
occur, and for the second order phase transition it allows tuning
the values of critical exponents \cite{pan27}. An interesting
extension was done in \cite{pan28} by constructing general models
for holographic superconductivity. Rich phenomena in the phase
transition were also found for the holographic superconductors in
Einstein-Gauss-Bonnet gravity where the Gauss-Bonnet coupling can
play the role in determining the order of phase transition and
critical exponents in the second-order phase transition
\cite{pan30}.

Besides the bulk AdS black hole spacetime, recently a
superconducting phase dual to an AdS soliton configuration was
studied in \cite{pan127}. The AdS soliton is a gravitational
configuration which has lower energy than the AdS space in the
Poincar\'{e} coordinates, but has the same boundary topology as the
Ricci flat black hole and the AdS space in the Poincar\'{e}
coordinates \cite{pan124}. There is a first order phase transition
called Hawking-Page phase transition between the Ricci flat AdS
black hole and the AdS soliton \cite{pan125}. The signature of this
phase transition was shown up in the quasi-normal modes spectrum
\cite{pan126}. The Hawking-Page phase transition between Ricci flat
black holes and deformed AdS soliton in the Gauss-Bonnet gravity was
discussed in \cite{pan123}. In \cite{pan127}, another phase
transition was disclosed in the AdS soliton configuration. It was
shown that if one adds a chemical potential $\mu$ to the AdS
soliton, there is a second order phase transition at a critical
value $\mu_c$, beyond which the charge scalar field turns on, even at
zero temperature. This is an insulator/superconductor transition.
The analysis in \cite{pan127} was performed in the limit of the
probe approximation, where the backreaction of the matter fields is
not taken into account. Considering the backreaction of the matter
fields on the soliton geometry, a new phase boundary in which a
first order phase transition was observed for the same chemical
potential when the backreaction is strong enough \cite{Gary
T.Horowitz-3}. The various phase transitions of gravity duals of
superfluid/fluid/insulator in (2+1)-dimensions have been
investigated in \cite{brihaye}.

In this work we will extend the discussion on the basic holographic
insulator/superconductor transition in the AdS soliton background by
generalizing the spontaneous breaking of a global U(1) symmetry to
occur via St$\ddot{u}$ckelberg mechanism. We will construct the
soliton solutions with backreaction, examine the effects of the
backreaction on the condensation of the scalar hair in the
generalized St$\ddot{u}$ckelberg Lagrangian and disclose rich
physics in various phase transitions. In addition to the AdS soliton
background, we will also examine the property of the phase
transition in AdS black
hole background in the generalized St$\ddot{u}$ckelberg formalism.

\section{various phase transitions in the AdS soliton background}

\subsection{The bulk equations}

In this section, we study in detail the formation of the scalar hair
in the 5-dimensional AdS soliton spacetime. The generalized action
containing a U(1) gauge field and the scalar field coupled via a
generalized St$\ddot{u}$ckelberg Lagrangian reads
\cite{pan26,pan27,pan30}
\begin{eqnarray}\label{lagrange-1}
S&=&\int
d^5x\sqrt{-g}\left[\left(R+\frac{12}{l^{2}}\right)+L_{matter}\right],
\end{eqnarray}
where $l$ is the AdS radius which will be taken to be unity in the
following discussion. $L_{matter}$ is the generalized
St$\ddot{u}$ckelberg Lagrangian
\begin{eqnarray}\label{lagrange-3}
L_{matter}&=-\frac{1}{4}F^{\mu\nu}F_{\mu\nu}-(\partial{\psi})^{2}
 -m^{2}\mid{{\psi}}\mid^{2} -(\psi^{2}+q^{2}c_{4}\psi^{4})(\partial{p}-qA)^{2},
\end{eqnarray}
where $\psi(r)$ and $A_{\mu}$ are the scalar and Maxwell fields
coupled to gravity. $c_{4}$ is a model parameter. Setting  $qA=\tilde{A}$
and considering the gauge symmetry
$\tilde{A}_{\mu}\rightarrow~\tilde{A}_{\mu}+\partial\Lambda$~and~$p\rightarrow~p+\Lambda$,
we can fix the gauge $p = 0$ by using the gauge freedom.

We are interested in including the backreaction so we choose the
metric ansatz
\begin{eqnarray}\label{metric}
ds^{2}&=&r^{2}\left[e^{A(r)}B(r)d\eta^{2}+dx^{2}+dy^{2}-e^{C(r)}dt^{2}\right]+\frac{dr^{2}}{r^{2}B(r)}~.
\end{eqnarray}
We require that $B$ vanishes at some radius $r_s$, which corresponds
to the tip of the soliton. As in the usual AdS soliton, smoothness
at the tip requires that $\eta$ be periodic with the period
\begin{eqnarray}\label{period}
\gamma=\frac{4\pi e^{-A(r_{s})/2}}{r_{s}^{2}B'(r_{s})}.
\end{eqnarray}

Choosing the electromagnetic field and the scalar field as
\begin{eqnarray}\label{MF}
\psi~=~\psi(r)~,~~~~A~=~\phi(r) dt~,
\end{eqnarray}
we can obtain the equations of motion of matter fields
\begin{eqnarray}\label{psi}
\psi''+\left(\frac{5}{r}+\frac{A'}{2}+\frac{B'}{B}+\frac{C'}{2}\right)\psi'
+\frac{q^{2}\phi^{2}e^{-C}}{r^{4}B}(\psi
+2q^{2}c_{4}\psi^{3})-\frac{m^{2}}{r^{2}B}\psi=0,
\end{eqnarray}
\begin{eqnarray}\label{phi}
\phi''+\left(\frac{3}{r}+\frac{A'}{2}+\frac{B'}{B}-\frac{C'}{2}\right)\phi'+
\frac{2q^{2}\phi}{r^{2}B}(\psi^{2}+q^{2}c_{4}\psi^{4})=0.
\end{eqnarray}

The nontrivial components, such as tt, rr, xx, yy, $\eta\eta$
components in Einstein equations  are not all independent. From the
xx-tt and $\eta\eta$+tt-xx components, we have
\begin{eqnarray}\label{metric-1}
C''+\frac{1}{2}C'^{2}+\left(\frac{5}{r}+\frac{A'}{2}+\frac{B'}{B}\right)C'-\left[\phi'^{2}+
\frac{2q^{2}\phi^{2}}{r^{2}B}(\psi^{2}+q^{2}c_{4}\psi^{4})\right]\frac{e^{-C}}{r^{2}}=0,
\end{eqnarray}
\begin{eqnarray}\label{metric-2}
B'\left(\frac{3}{r}-\frac{C'}{2}\right)+B\left(\psi'^{2}-\frac{1}{2}A'C'+\frac{e^{-C}\phi'^{2}}{2r^{2}}+
\frac{12}{r^{2}}\right)+\frac{q^{2}\phi^{2}e^{-C}}{r^{4}}(\psi^{2}+q^{2}c_{4}\psi^{4})+
\frac{1}{r^{2}}(m^{2}\psi^{2}-12)=0.
\end{eqnarray}
Substracting the $\eta\eta$ component from the rr equation, we get
\begin{eqnarray}\label{metric-3}
A'=\frac{2r^{2}C''+r^{2}C'^{2}+4rC'+4r^{2}\psi'^{2}-2e^{-C}\phi'^{2}}{r(6+rC')}.
\end{eqnarray}

We need to integrate these equations from the tip of the soliton out
to the infinity. Since the equations are coupled and nonlinear, we
have to count on the numerical calculation.  At the tip $r_{s}$, we
can do the expansion
\begin{eqnarray}\label{boudary}
\psi(r)=a+b(r-r_{s})+c(r-r_{s})^{2}+\cdot\cdot\cdot,\\
\phi(r)=\alpha+\beta(r-r_{s})+\gamma(r-r_{s})^{2}+\cdot\cdot\cdot,\\
B(r)=u(r-r_{s})+\cdot\cdot\cdot,\\
C(r)=v+w(r-r_{s})+\cdot\cdot\cdot,\\
A(r)=A_{0}+A_{1}(r-r_{s})+\cdot\cdot\cdot.
\end{eqnarray}
Putting these Taylor expansions into equations, we get four
independent parameters: $a, A_{0}, \alpha, v$. At
$r\rightarrow\infty$, after choosing $m^{2}=-\frac{15}{4}$ above the
BF bound for $AdS_{5}$: $m^{2}_{BF}=-\frac{(D-1)^{2}}{4}=-4$
\cite{P.Breitenlohner}, the scalar and Maxwell equations have the
form
\begin{eqnarray}\label{inf}
\psi=\frac{\psi_{-}}{r^{3/2}}+\frac{\psi_{+}}{r^{5/2}}+\cdot\cdot\cdot,\
\phi=\mu-\frac{\rho}{2r^{2}}+\cdot\cdot\cdot,
\end{eqnarray}
and parameters $C(\infty)=0, A(\infty)=0, B(\infty)=1$. At our
chosen value of $m^2$, both $\psi_{-}$ or $\psi_{+}$ are
normalizable. We fix $\psi_{-}=0$ in our following discussion.  For
given values of $a$, we can obtain the required numerical solutions
from the differential equations above.

Using the scaling symmetry
\begin{eqnarray}\label{symmetry}
r \rightarrow ar,~~~~~~~~(t,x,y,\eta)\rightarrow
~(t,x,y,\eta)/a,~~~~~~~\phi\rightarrow a\phi,
\end{eqnarray}
we can set $r_{s}=1$. After getting the solutions, these symmetry
can be used  again to scale $\gamma=\pi$. Other quantities and
operators are scaled as: $\gamma$ $\rightarrow$ $\frac{1}{a}$
$\gamma$, ~$\mu$ $\rightarrow$ a$\mu$, ~$\rho$ $\rightarrow$
$a^{3}$$\rho$, ~$\psi_{+}^{5/2}$$\rightarrow$a$\psi_{+}^{5/2}$.

\subsection{Effects on the phase transition}

It was shown  in \cite{liu26,pan127} that AdS soliton experiences a
second order phase transition, the insulator/superconductor phase
transition, around the critical chemical potential $\mu_{c}$. But
these results were got in the probe limit which was justified for
large $q$. Taking  the backreaction into account, it was observed
that when $q$ is reduced to $q\approx 1$, the second order phase
transition will give way to the first order phase
transition\cite{Gary T.Horowitz-3}.

It is of interest to see how the generalized St$\ddot{u}$ckelberg
formalism plays the role in the phase transition in the AdS soliton
background. When  $c_4=0$, our result goes back to that disclosed in
\cite{Gary T.Horowitz-3} that the first order phase transition
appears only when $q\sim 1$ and for $q>1$ there is only the second
order phase transition. For $q>1$, we need to turn on $c_4$ to
accommodate the first order phase transition. For the chosen
strength of the backreaction with $q>1$, we can find a threshold
value of $c_4$. Above this threshold $c_4$, the condensate operator
together with the charge density $\rho$ do not have a monotonic
behavior, which indicates the appearance of the first order phase
transition. When we increase $q$, we find that the threshold
$\bar{c}_{4}$ becomes bigger. We show the numerical behaviors of the
condensate and the charge density with the change of the chemical
potential in Fig.1 and Fig.2.

\begin{figure}[h]
\includegraphics[width=2.8in,height=2in]{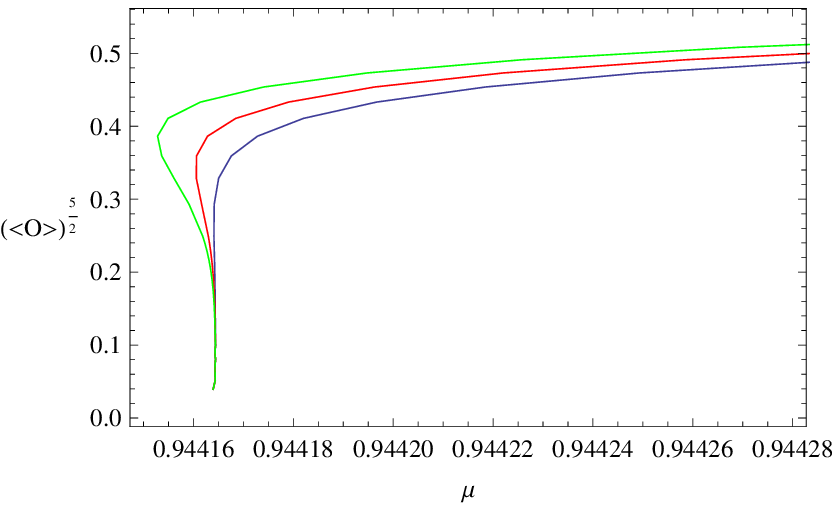}\
\includegraphics[width=2.8in,height=2in]{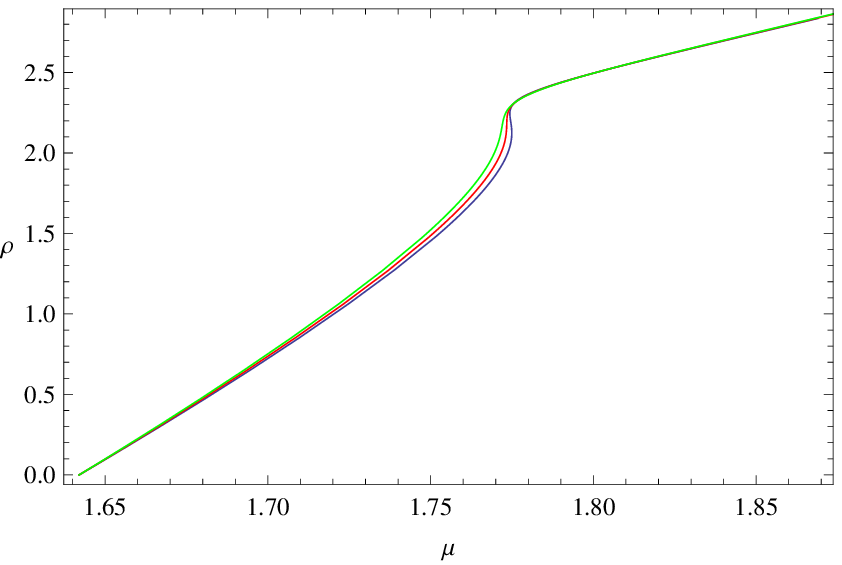}\
\caption{\label{fig1}(Color online) The left panel shows the value of the condensate
$<O>$ as a function of chemical potential $\mu$ when $q=2$ with $c_{4}=0.471$ (blue), $c_{4}=0.473$ (red),
$c_{4}=0.475$ (green), respectively. The right panel shows the charge density $\rho$ as a
function of chemical potential $\mu$ with q$=$1.15. $c_{4}$
decreases from left to right, i.e., $c_{4}=0.11$ (green),
$c_{4}=0.105$ (red) and $c_{4}=0.1$ (blue). }
\end{figure}

We observe the influence of the backreaction on the scalar
condensation in the AdS soliton background. For choosing the same
value of the scalar mass, qualitative features occur as we vary the
backreaction of  the matter field on the AdS soliton background. As
we increase $q$, the condensation gap becomes smaller, which means
that the scalar hair can be formed easier in the more weakly
backreacted background. The gap is not sensitive to the model
parameter $c_4$.

It is interesting to ask what is the upper bound on $\bar{c}_{4}$
when $q$ grows to infinity, namely when  the backreaction
disappears. In order to answer this question, we set
$q\psi=\tilde{\psi}$, $q\phi=\tilde{\phi}$,
$\frac{1}{q^{2}}=\lambda$, so that
\begin{eqnarray}\label{psi-1}
\tilde{\psi}''+\left(\frac{5}{r}+\frac{A'}{2}+\frac{B'}{B}+\frac{C'}{2}\right)\tilde{\psi}'+
\frac{\tilde{\phi}^{2}e^{-C}}{r^{4}B}\left(\tilde{\psi}+2c_{4}\tilde{\psi}^{3}\right)-
\frac{m^{2}}{r^{2}B}\tilde{\psi}=0,
\end{eqnarray}
\begin{eqnarray}\label{phi-1}
\tilde{\phi}''+\left(\frac{3}{r}+\frac{A'}{2}+\frac{B'}{B}-\frac{C'}{2}\right)\tilde{\phi}'-
\frac{2\tilde{\phi}}{r^{2}B}\left(\tilde{\psi}^{2}+c_{4}\tilde{\psi}^{4}\right)=0,
\end{eqnarray}
\begin{eqnarray}\label{C-2}
C''+\frac{1}{2}C'^{2}+\left(\frac{5}{r}+\frac{A'}{2}+\frac{B'}{B}\right)C'-\lambda\left[\tilde{\phi}'^{2}+
\frac{2\tilde{\phi^{2}}}{r^{2}B}\left(\tilde{\psi}^{2}+c_{4}\tilde{\psi}^{4}\right)\right]\frac{e^{-C}}{r^{2}}=0,
\end{eqnarray}
\begin{eqnarray}\label{B-2}
B'\left(\frac{3}{r}-\frac{C'}{2}\right)+B\left(\lambda\tilde{\psi}'^{2}-\frac{1}{2}A'C'+\lambda\frac{e^{-C}
\tilde{\phi}'^{2}}{2r^{2}}+\frac{12}{r^{2}}\right)+\lambda\frac{\tilde{\phi}^{2}e^{-C}}{r^{4}}
\left(\tilde{\psi}^{2}+c_{4}\tilde{\psi}^{4}\right)+\frac{1}{r^{2}}\left(\lambda
m^{2}\tilde{\psi}^{2}-12\right)=0,
\end{eqnarray}
\begin{eqnarray}\label{A-2}
A'=\frac{2r^{2}C''+r^{2}C'^{2}+4rC'+4r^{2}\lambda\tilde{\psi}'^{2}-2e^{-C}\lambda\tilde{\phi}'^{2}}{r(6+rC')}.
\end{eqnarray}
$q\rightarrow\infty$ corresponds to $\lambda\rightarrow0$ which
reduces all equations above to the probe limit. Solving the above
equations numerically, we find $\bar{c}_{4}=0.57$ when
$q\rightarrow\infty$, which means that even in the probe limit, when
$c_4$ is above this threshold value, the first order phase
transition can substitute the second order phase transition to
happen in the AdS soliton background. This is qualitatively similar
to the findings in the AdS black hole configurations
\cite{pan27,pan30}.  For clarity, in Table \ref{ADS soliton} we list
the values of the threshold $\bar{c}_{4}$ for different strength of
the backreaction in the AdS soliton background.

\begin{table}[ht]
\begin{center}
\caption{\label{ADS soliton} The threshold value $\bar{c}_{4}$ which
can accommodate the first order phase transition for different
strength of the backreaction in the AdS soliton background.}
\begin{tabular}{|c|c|c|c|c|c|c|}
 \hline
q & 1.0 & 1.5  & 1.7 &2.0&5.0&$\infty$\\
\hline
$~\bar{c}_{4}$~&~0.000~&~0.230~&~0.450~&~0.473~&~0.555~&~0.570~ \\
\hline
\end{tabular}
\end{center}
\end{table}

In the above discussion we fixed the strength of the backreaction
and examined the threshold  of $c_4$ in the  generalized
St$\ddot{u}$ckelberg formalism to change the order of the phase
transition from the second to the first. We have also examined the effects of the change of the strength
of the backreaction on the order of the phase transition for
the chosen $c_4$. We find that for the chosen nonzero $c_4$,
the condensate exhibits non-monotonic behavior for smaller
$q$. This means that the first order of the phase transition can
appear easily when the backreaction is stronger for the fixed $c_4$.

\section{Various phase transitions in the AdS black hole background}

\subsection{The bulk equations}

Now we turn our discussion to the AdS black hole background. To
include the backreaction, we consider the metric ansatz
\begin{eqnarray}\label{metric}
ds^{2}&=&-g(r)e^{-\chi(r)}dt^{2}+\frac{dr^{2}}{g(r)}+r^{2}(dx^{2}+dy^{2}+dz^{2})~.
\end{eqnarray}
The event horizon $r_{H}$ satisfies $g(r_{H})=0$ and the Hawking
temperature is expressed as
$T=\frac{g'(r_{H})e^{-\chi(r_{H})/2}}{4\pi}$. Using the symmetry
\begin{eqnarray}\label{boudary-1}
r \rightarrow ar, (t,x,y,z) \rightarrow (t,x,y,z)/a, g \rightarrow
a^{2}g, \phi \rightarrow a\phi,
\end{eqnarray}
we can set the horizon $r_{H}=1$.

Assuming the matter fields in the form of (\ref{MF}) , we get the
equations of motion
\begin{eqnarray}\label{BH-1}
\psi''+\left(\frac{3}{r}-\frac{\chi'}{2}+\frac{g'}{g}\right)\psi'+\frac{q^{2}\phi^{2}e^{\chi}}
{g^{2}}\left(\psi+2q^{2}c_{4}\psi^{3}\right)-\frac{m^{2}}{g}\psi=0,
\end{eqnarray}
\begin{eqnarray}\label{BH-2}
\phi''+\left(\frac{3}{r}+\frac{\chi'}{2}\right)\phi'-\frac{2q^{2}\phi}{g}\left(\psi^{2}+q^{2}c_{4}\psi^{4}\right)=0,
\end{eqnarray}
\begin{eqnarray}\label{BH-3}
\chi'+\frac{2r\psi'}{3}+\frac{2rq^{2}\phi^{2}e^{\chi}}{3g^{2}}\left(\psi^{2}+q^{2}c_{4}\psi^{4}\right)=0,
\end{eqnarray}
\begin{eqnarray}\label{BH-4}
g'+\left(\frac{2}{r}-\frac{\chi'}{2}\right)g+\frac{re^{\chi}\phi'^{2}}{6}+\frac{m^{2}r\psi^{2}}{3}-4r=0.
\end{eqnarray}
For the Maxwell field which is regular at the horizon, we can choose
$\phi(r_{H})=0$. The solution around the horizon can be expanded as

\begin{eqnarray}\label{boudary-1}
\psi(r)=a'+b'(r-r_{H})+c'(r-r_{H})^{2}+\cdot\cdot\cdot,~~~~~\\
\phi(r)=\alpha'(r-r_{H})+\beta'(r-r_{H})^{2}+\cdot\cdot\cdot,~~~~~~~~~\\
g(r)=u'(r-r_{H})+\cdot\cdot\cdot,~~~~~~~~~~~~~~~~~~~~~~~~~~~~\\
\chi(r)=v'+w'(r-r_{H})+\cdot\cdot\cdot.~~~~~~~~~~~~~~~~~~~~~~~
\end{eqnarray}

Substituting them into the equations of motion, we find that $a',
u', v'$ are independent. For the mass $m^{2}=-\frac{15}{4}$, the
scalar and Maxwell fields have the same asymptotic behavior just as
shown in Eq. (\ref{inf}). For simplicity, we will choose
$a'$ as the parameter to adjust the solution to satisfy $\psi_{-}=0$
and $\chi(\infty)=0$. We can integrate these equations numerically
from the horizon out to infinity.

\subsection{Effects on the phase transition}

After solving the differential equations numerically, we can examine
the phase transition in the AdS black hole background. When $c_4=0$,
the phase transition always remains second order in the AdS black
hole background even when $q<1$, which supports the observation in
\cite{Gary T.Horowitz-3}. The second order phase transition will
give way to the first order only when we turn on $c_4$ in the
generalized St$\ddot{u}$ckelberg formalism. In the probe limit when
$q\rightarrow\infty$, we obtain $\bar{c}_{4}=2.28$. Below this
$\bar{c}_{4}$, the phase transition is of the second order, while
the first order phase transition is allowed when $c_4>2.28$. When
the backreaction of the matter fields to the AdS black hole
background is not strong enough, the behavior of the condensation
due to the change of $c_4$ was shown in Fig.4. The condensate
operator appears a non-monotonic behavior which indicates the
appearance of the first order phase transition. We find that when
$q$ is not small enough, with the decrease of the value of $q$, the
threshold value of $c_4$ to allow the first order phase transition
to happen becomes smaller.

\begin{figure}[h]
\includegraphics[width=2.8in,height=2in]{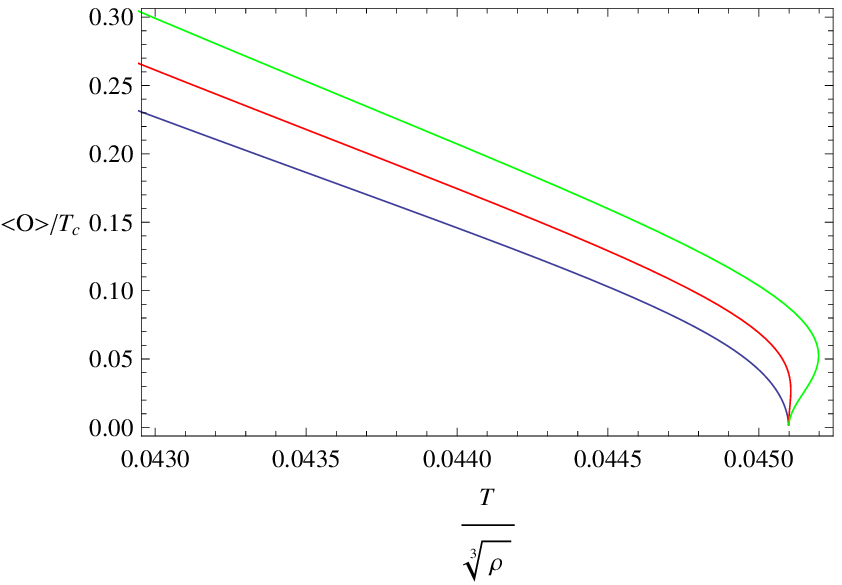}
\includegraphics[width=2.8in,height=2in]{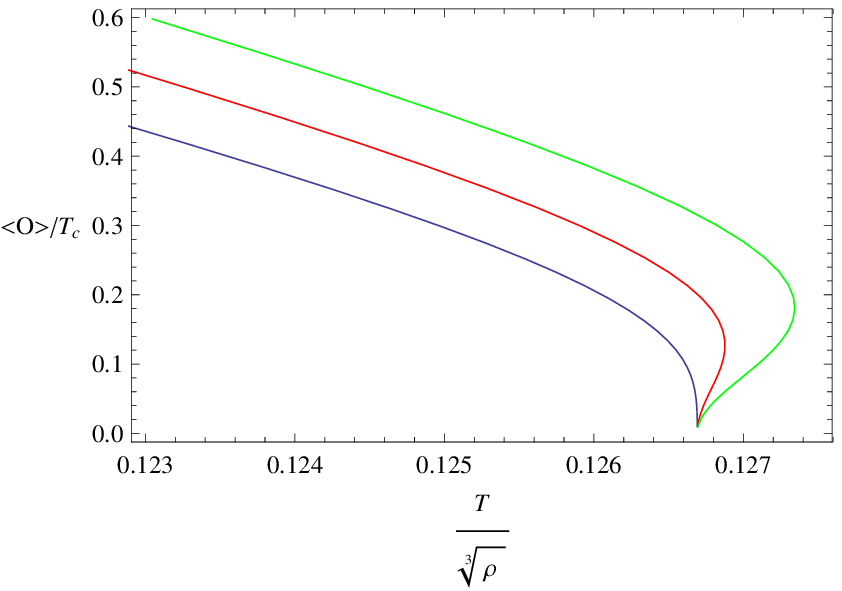}
\caption{\label{fig5}(Color online) The value of the condensate
$<O>$ as a function of the temperature $T$. The
left column represents the case when $q=1$ and the three lines from
left to right correspond to the increase of $c_{4}$, i.e.,
$c_{4}=0.74$ (blue), $c_{4}=0.75$ (red), $c_{4}=0.76$ (green), the right one is for $q=1.5$ with $c_4$ changing
($c_{4}=0.4$ (blue), $c_{4}=0.425$ (red), $c_{4}=0.45$ (green)).}
\end{figure}

However the threshold $\bar{c}_{4}$ will not decrease monotonically
with  the decrease of $q$ (the increase of the backreaction). When
the backreaction is strong enough for some certain small value of
$q$, with the decrease of $q$, the threshold $\bar{c}_{4}$ will
increase to allow the first order phase transition to happen. For
different chosen values of $q$, the thresholds $\bar{c}_{4}$ to
accommodate the first order phase transition are listed in table II.

\begin{table}[ht]
\begin{center}
\caption{\label{ADS BH} The threshold value $\bar{c}_{4}$ which can
accommodate the first order phase transition for different strength
of the backreaction in the AdS black hole background.}
\begin{tabular}{|c|c|c|c|c|c|c|}
 \hline
q & 1.0 & 1.5  & 2.0 &5.0&10.0&$\infty$\\
\hline
$~\bar{c}_{4}$~&~0.75~&~0.43~&~0.53~&~1.55~&~2.05~&~2.28~ \\
\hline
\end{tabular}
\end{center}
\end{table}

From the equations of motion (\ref{BH-1})-(\ref{BH-4}), we see that
the $c_4$ is always related to $q^4$. This implies that in the
generalized St$\ddot{u}$ckelberg formalism in order to show the
effect of $c_4$ when $q$ approaches zero, $c_4$ must tend to
infinity. To check this, more careful numerical calculations are
needed. However, when $q<1$, our numerical calculation becomes not efficient and more time
consuming with the decrease of $q$.

\section{Conclusions and discussions}

We have studied the possibility to allow various phase transitions
to happen in the AdS soliton configuration. In the probe limit, it
was argued that only second order phase transition can happen in the
AdS soliton background [24,35]. Considering the backreaction of the
matter field to the background spacetime, it was found that when the
effect of the backreaction is strong enough, for example with $q\sim 1$, the second order phase transition will give way to the first order phase
transition between the insulator and superconductor in
the AdS soliton configuration [41].  It is of interest to study the phase transition in the generalized St$\ddot{u}$ckelberg formalism, which has been applied in the AdS black hole backgrounds [31,32,34]. Applying the
generalized St$\ddot{u}$ckelberg formalism to the AdS soliton configuration, we find that there is always a threshold which can
change the phase transition from the second order to the first
order. When the backreaction of the matter field becomes weaker, the
threshold in the generalized St$\ddot{u}$ckelberg formalism to
accommodate the first order phase transition is bigger. But
there is an upper limit of this threshold, above this upper limit
the first order phase transition can happen even in the probe limit
in the AdS soliton configuration. In the AdS soliton case, the transition of the second order phase transition into the first order is natural as there is already Hawking-page transition which is the first order.

We have extended our discussion to the AdS black hole background. It
was argued in \cite{Gary T.Horowitz-3} that in the AdS black hole
spacetime, no matter how strong the backreaction is considered, even with $q<1$,
there only exists the second order phase transition between the
conductor and superconductor. In  the generalized
St$\ddot{u}$ckelberg formalism, we observed rich physics in the
phase transition in the AdS black hole configuration.
When the
backreaction is weak, there is a threshold and above which the first
order phase transition can happen. The threshold value becomes
smaller when the backreaction becomes bigger. However when the
backreaction is very strong, the first order phase transition will
be hard to happen in the AdS black hole configuration. For the very strong backreaction, the relation
between the threshold and the backreaction is different from that
for the weak backreaction case, the threshold increases when the
backreaction becomes much stronger. Counting on the big effect of the $\Psi^4$ term in the generalized
St$\ddot{u}$ckelberg formalism, we found that the first order phase transition can be accommodated in the strong backreaction situation, for example with $q<1$.

When the backreaction is very strong with $q$ much smaller than unity, we argued that
the threshold value to accommodate the first order
phase transition will experience a fast grow. This argument still needs further accurate numerical check. Our
present numerical code becomes inefficient and very time consuming when the
backreaction $q<1$. More precise calculations are called
for.

In \cite{xx} it was argued that transforming the Abelian Higgs model in three dimensions (i.e. the Ginzburg-Landau theory) to a disorder field theory, there is a critical point where the second order phase transition changes to the first order, which happens for a specific range of the Ginzburg-Landau parameter. Our observation of the change of the order of the phase transition in the AdS black hole background gives a very similar example in the gravity side. This shows a further example of the correspondence between the properties in AdS spacetime and the condensed matter physics.

\begin{acknowledgments}
This work was partially supported by NNSF of China.

\end{acknowledgments}


\begin{thebibliography}{99}


\bibitem{Maldacena}
J.M. Maldacena, Adv. Theor. Math. Phys. {\bf 2}, 231 (1998).

\bibitem{S.S.Gubser-1}
S.S. Gubser, I.R. Klebanov, and A.M. Polyakov, Phys. Lett. B {\bf
428}, 105 (1998).

\bibitem{E.Witten}
E. Witten, Adv. Theor. Math. Phys. {\bf 2}, 253 (1998).

\bibitem{liu4}
S.A. Hartnoll, Class. Quant. Grav. {\bf 26}, 224002 (2009).

\bibitem{liu5}
C.P. Herzog, J. Phys. A {\bf 42}, 343001 (2009).

\bibitem{liu6}
G.T. Horowitz, arXiv:1002.1722 [hep-th].

\bibitem{liu7}
G.T. Horowitz and M.M. Roberts, Phys. Rev. D {\bf 78}, 126008
(2008).

\bibitem{liu8}
E. Nakano, W.Y. Wen, Phys. Rev. D {\bf 78}, 046004 (2008).

\bibitem{liu9}
I. Amado, M. Kaminski, and K. Landsteiner, J. High Energy Phys. {\bf
0905}, 021 (2009).

\bibitem{liu10}
G. Koutsoumbas, E. Papantonopoulos, and G. Siopsis, J. High Energy
Phys. {\bf 0907}, 026 (2009).

\bibitem{liu11}
O.C. Umeh, J. High Energy Phys. {\bf 0908}, 062 (2009).

\bibitem{liu12}
H.B. Zeng, Z.Y. Fan, and Z.Z. Ren, Phys. Rev. D {\bf 80}, 066001
(2009).

\bibitem{liu13}
J. Sonner, Phys. Rev. D {\bf 80}, 084031 (2009).

\bibitem{liu14}
S.S. Gubser, C.P. Herzog, S.S. Pufu, and T. Tesileanu, Phys. Rev.
Lett. {\bf 103}, 141601 (2009).

\bibitem{liu15}
J.P. Gauntlett, J. Sonner, and T. Wiseman, Phys. Rev. Lett. {\bf
103}, 151601 (2009).

\bibitem{liu16}
R.G. Cai and H.Q. Zhang, Phys. Rev. D {\bf 81}, 066003 (2010).

\bibitem{liu17}
J.L. Jing and S.B. Chen, Phys. Lett. B {\bf 686}, 68 (2010).



\bibitem{liu20}
C.P. Herzog, Phys. Rev. D {\bf 81}, 126009 (2010).

\bibitem{liu21}
S.B. Chen, L.C. Wang, C.K. Ding, and J.L. Jing, Nucl. Phys. B {\bf
836}, 222 (2010).

\bibitem{liu22}
R.A. Konoplya and A. Zhidenko, Phys. Lett. B {\bf 686}, 199 (2010).

\bibitem{liu23}
G. Siopsis and J. Therrien, J. High Energy Phys. {\bf 1005}, 013
(2010).

\bibitem{liu24}
K. Maeda, M. Natsuume, and T. Okamura, Phys. Rev. D {\bf 79}, 126004
(2009).

\bibitem{liu25}
R. Gregory, S. Kanno, and J. Soda, J. High Energy Phys. {\bf 0910},
010 (2009).

\bibitem{liu26}
Q.Y. Pan, B. Wang, E. Papantonopoulos, J. Oliveira, and A.B. Pavan,
Phys. Rev. D {\bf 81}, 106007 (2010).

\bibitem{liu27}
X.H. Ge, B. Wang, S.F. Wu, and G.H. Yang, J. High Energy Phys. {\bf
1008}, 108 (2010).

\bibitem{liu28}
X. He, B. Wang, R.G. Cai, and C.Y. Lin, Phys. Lett. B {\bf 688}, 230
(2010).

\bibitem{liu29}
R.G. Cai, Z.X. Nie, B. Wang, and H.Q. Zhang, arXiv:1005.1233
[gr-qc].


\bibitem{liu31}
Y. Brihaye and B. Hartmann, Phys. Rev. D {\bf 81}, 126008 (2010).

\bibitem{liu32}
S.S. Gubser, Phys. Rev. D {\bf 78}, 065034 (2008).

\bibitem{liu33}
S.A. Hartnoll, C.P. Herzog, and G.T. Horowitz, Phys. Rev. Lett. {\bf
101}, 031601 (2008).


\bibitem{pan26}
S. Franco, A.M. Garcia-Garcia, and D. Rodriguez-Gomez, J. High
Energy Phys. {\bf 1004}, 092 (2010).

\bibitem{pan27}
S. Franco, A.M. Garcia-Garcia, and D. Rodriguez-Gomez, Phys. Rev. D
{\bf 81}, 041901(R) (2010).

\bibitem{pan28}
F. Aprile and J.G. Russo, Phys. Rev. D {\bf 81}, 026009 (2010).

\bibitem{pan30}
Q.Y. Pan and B. Wang, Phys. Lett. B {\bf 693}, 159 (2010).

\bibitem{pan127}
T. Nishioka, S. Ryu, and T. Takayanagi, J. High Energy Phys. {\bf
1003}, 131 (2010).

\bibitem{pan124}
G.T. Horowitz and R.C. Myers, Phys. Rev. D {\bf 59}, 026005 (1998).

\bibitem{pan125}
S. Surya, K. Schleich, and D.M. Witt, Phys. Rev. Lett. {\bf 86},
5231 (2001).

\bibitem{pan126}
J. Shen, B. Wang, R. K. Su, C.Y. Lin, and R.G. Cai, J. High Energy
Phys. {\bf 0707}, 037 (2007).

\bibitem{pan123}
R.G. Cai, S.P. Kim, and B. Wang, Phys. Rev. D {\bf 76}, 024011
(2007).

\bibitem{Gary T.Horowitz-3}
G.T. Horowitz and B. Way, J. High Energy Phys. {\bf 1011}, 011
(2010).

\bibitem{brihaye}
Y. Brihaye and B. Hartmann, arXiv:1101.5708 [hep-th].

\bibitem{P.Breitenlohner}
P. Breitenlohner and D.Z. Freedman, Phys. Lett. B {\bf 115}, 197
(1982).

\bibitem{xx} H. Kleinert, Lett. Nuovo Cimento, 35, (1982) 405.


\end{thebibliography}
\end{document}